\documentclass{phb-proc4-auth}

\usepackage{graphicx}
\usepackage{amssymb}
\usepackage{amsmath}
\newcommand{\cal}{\mathcal}
\newcommand{\tth}{$^{\text{th}}$\ }
\newcommand{\pdagger}{{\phantom{\dagger}}}

\newcommand{\PT}{_{\text{PT}}}
\newcommand{\refq}[1]{(\ref{eq:#1})}
\newcommand{\reff}[1]{Fig.~\ref{fig:#1}}

\begin{document}
\begin{frontmatter}


\journal{SCES '04}


\title{
Ground state of the frustrated Hubbard model within DMFT:\\
energetics of Mott insulator and metal from ePT and QMC}


\author[Mainz]{N.~Bl\"umer\corauthref{1}}
\author[Marburg]{E.~Kalinowski}

%
 
\address[Mainz]{Institut f\"ur Physik, Johannes-Gutenberg-Universit\"at, 55099 Mainz, Germany}
\address[Marburg]{FB Physik, Philipps Universit\"at, 35032 Marburg, Germany}

%
%

\corauth[1]{
Nils.Bluemer@uni-mainz.de}


\begin{abstract}
  We present a new method, ePT, for extrapolating few known
  coefficients of a perturbative expansion. Controlled by comparisons
  with numerically exact quantum Monte Carlo (QMC) results,
  $10^{\text{th}}$ order strong-coupling perturbation theory (PT) for
  the Hubbard model on the Bethe lattice is reliably extrapolated to
  infinite order. Within dynamical mean-field theory (DMFT), we obtain
  continuous estimates of energy $E$ and double occupancy $D$ with
  unprecedented precision ${\cal O}(10^{-5})$ for the Mott insulator
  above its stability edge $U_{c1}\approx 4.78$ as well as critical
  exponents.  In addition, we derive corresponding precise estimates
  for $E$ and $D$ in the metallic ground state from extensive
  low-temperature QMC simulations using a fit to weak-coupling PT
  while enforcing thermodynamic consistency.
\end{abstract}

%
%



\end{frontmatter}


The half-filled Hubbard model has been subject of intense
research; after clarifying the magnetic phases, DMFT studies have
lately focused on the paramagnetic Mott metal-insulator transition
observed in frustrated versions. In the following, we shortly
summarize recent work; for more complete and self-contained
presentations, see Ref. \cite{Bluemer04a} and upcoming publications.

Kato-Takahashi perturbation theory (PT) for the half-filled frustrated
Hubbard model
  \[
    \hat{H} = -\sum_{(ij),\sigma} t_{ij}
             \left(\hat{c}_{i\sigma}^\dagger \hat{c}_{j\sigma}^\pdagger
             + \hat{c}_{j\sigma}^\dagger \hat{c}_{i\sigma}^\pdagger\right) 
             + U \hat{D}
  \]
with $\hat D=\sum_{i}\hat n_{i\uparrow}\hat n_{i\downarrow}$,
$\hat{n}_{i\sigma}=\hat{c}_{i\sigma}^\dagger\hat{c}_{i\sigma}^\pdagger$,
 and 
 semi-elliptic ``Bethe'' density of states 
 ($W=4t$) yields \cite{Bluemer04a} the energy $E(U)=E\PT(U)+ {\cal
    O}(t^{12}\, U^{-11})$, where for $t=1$:
  \begin{equation}\label{eq:E_PT}
    E\PT(U)=-\frac{1}{2\, U} -\frac{1}{2\,U^3} -\frac{19}{8\, U^5} 
    -\frac{593}{32\, U^7} -\frac{23877}{128\, U^9}\,.
  \end{equation}
Both the 10\tth order (in $t$) expression \refq{E_PT} for $E\PT$ and
the resulting estimate $D\PT=dE\PT/dU$ for the double occupancy
are converged within ${\cal O}(10^{-5})$ for $U\ge 6$.
Similar precision can be reached at finite temperatures $T$ using
an advanced quantum Monte-Carlo scheme \cite{Bluemer04a}; due to
the almost negligible temperature dependence in the insulating phase
this also applies to the Mott insulating ground state.
As seen in \reff{PT_QMC},
  \begin{figure}
  \includegraphics[width=\columnwidth,clip=true]{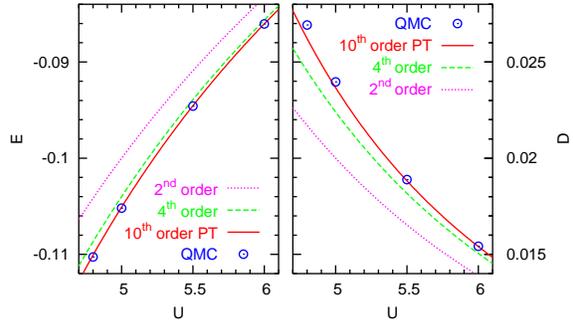}
  \caption{Energy $E$ and double occupancy $D$ of Mott insulator: 
    PT of various order (lines) in comparison with QMC (circles).}
  \label{fig:PT_QMC}
  \end{figure}
  both methods indeed agree very well for $U=6$ (and deviate below).

The idea of the extended perturbation theory (ePT) is to 
extrapolate coefficients in the PT series
$E\PT=\sum_{i=1}^\infty a_{2i}\, U^{1-2i}$
by fitting coefficient ratios $U_{c1}[2i]\equiv\sqrt{a_{2i+2}/a_{2i}}$ 
to $U_{c1}[n]\approx U_{c1} + u_1\,n^{-1} + u_2\, n^{-2}$.
As seen in \reff{PT_ext}, 
  \begin{figure}
  \includegraphics[width=\columnwidth,clip=true]{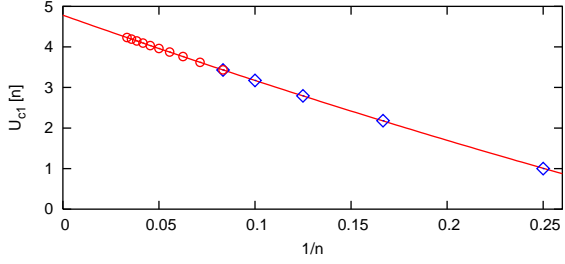}
  \caption{Construction of ePT: Extrapolation of PT values 
     for $U_c[n]$ (squares) defines coefficients for smaller $1/n$
    (circles).}
  \label{fig:PT_ext}
  \end{figure}
the convergence to this assumed asymptotic form is fast; consequently,
both the extrapolation to $U_{c1}\,=\,\lim_{i\to\infty} U_{c1}[2i]$
and the extraction of higher-order coefficients $U_{c1}[2i]$ are
reliable. We can further derive algebraic exponents: 
$a_n\,\propto\, n^{\tau} \,U_{c1}^n$,\,
$E_{\text{crit}}(U)\,\propto \,(U-U_{c1})^{{\tau}-1}$,\,
$D_{\text{crit}}(U)\,\propto\,(U-U_{c1})^{{\tau}-2}$. Analytic
and numerical arguments support \cite{Bluemer04a} $\tau=3.5$, 
$U_{c1}=4.78$. \reff{dE10dD10} 
  \begin{figure}
    \includegraphics[width=\columnwidth,clip=true]{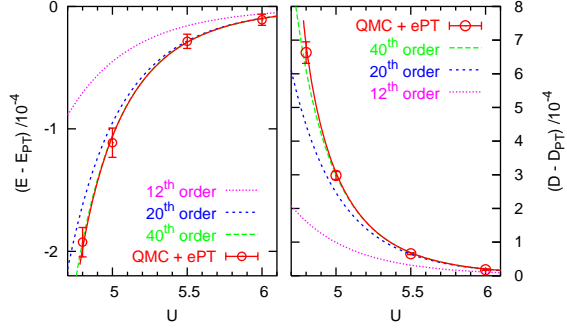}
  \caption{Mott insulator: Energy/double occupancy from QMC (circles), 
    ePT (solid line), and truncated ePT (dashed/dotted lines); all results 
   after subtracting 10\tth order PT.}
  \label{fig:dE10dD10}
  \end{figure}
shows the perfect agreement between ePT and QMC. These results mark
a breakthrough in accuracy for correlated electron systems; the ePT
curves (for which parameterizations are given in \cite{Bluemer04a})
should be/are used as authoritative benchmarks \cite{benchmarks}.
New evidence confirms our estimate of $U_{c1}=4.78$ which in turn
strongly supports our analysis of critical exponents (for which no
reliable comparisons exist).

Continuous ground state estimates are much more difficult to obtain from
finite-temperature methods (such as QMC) in the metallic phase due 
to the much stronger temperature dependencies (with divergent 
linear coefficient of specific heat for $U\to U_{c2}\approx 5.84$).
On the basis of careful extrapolations of data for $1/T=15,20,25,32,40$
(\reff{E_D_vsT2})
we have obtained such estimates with precision ${\cal O}(10^{-4})$
as shown in \reff{E0_D0_2opt2};
  \begin{figure}
\includegraphics[width=\columnwidth]{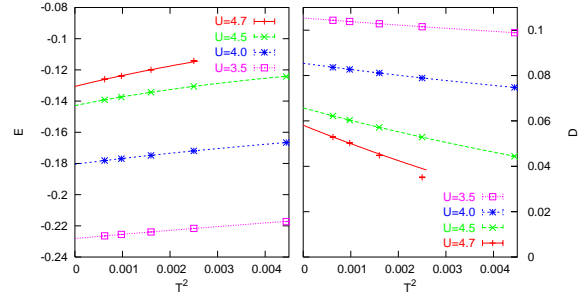}
\caption{Metal: QMC Energy/double occupancy (symbols) vs. squared temperature
$T$ and extrapolation $T\to 0$ (lines).}
  \label{fig:E_D_vsT2}
  \end{figure}
this accuracy suffices in order to discriminate between QMC
estimates (circles) and $2^{\text{nd}}$ order weak-coupling PT
(dotted lines). The continuous and thermodynamically consistent
fits (dashed lines) to QMC in \reff{E0_D0_2opt2} 
  \begin{figure}
\includegraphics[width=\columnwidth]{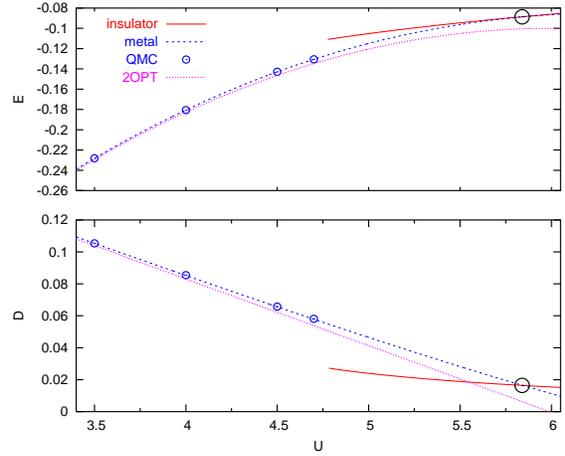}
\caption{Ground state of frustrated Hubbard model within DMFT:
Energy $E$ and double occupancy $D$ of metal and insulator match
at critical point $U_{c2}\approx 5.84$ (circle). $2^{\text{nd}}$ order
weak-coupling PT (dotted lines) shown for comparison.}
  \label{fig:E0_D0_2opt2}
  \end{figure}
  include the well-known $2^{\text{nd}}$ order term and fits to few
  higher even order coefficients under the constraint of matching the
  ePT estimates for the insulator at $U=U_{c2}\approx 5.84$ (circles).
  Note that our implicit estimate of the (weak-coupling) 4\tth order
  coefficient deviates strongly (and has opposite sign of) the
  analytic result found in the literature \cite{Mahlert03}.  Explicit
  parameterizations are deferred to a full-length publication which
  will also discuss additional observables implicit in our derivations
  (such as the linear coefficient of the specific heat).

%
%

%
%
%
%



\vspace{-1em}
\begin{thebibliography}{00}

\bibitem{Bluemer04a}
N.~Bl\"umer and E. Kalinowski, 
Preprint cond-mat/0404568 (2004) and references therein.
\bibitem{benchmarks} 
M.\ Feldbacher, K.\ Held, F.\ F.\ Assaad,
Preprint cond-mat/0405408 (2004);
S.\ Nishimoto, F.\ Gebhard, E.\ Jeckelmann,
Preprint cond-mat/0406666 (2004).
\bibitem{Mahlert03}
F. Gebhard, E. Jeckelmann, S. Mahlert, S. Nishimoto, R. M.~Noack,
Eur.\ Phys.\ J.\ B {\bf 36}, 491 (2003).
\end{thebibliography}
\end{document}